\documentclass[10pt]{article}
\usepackage{preamble, spconf, amsmath, graphicx, hyperref}

\usepackage{enumitem}
\setlist{nosep, leftmargin=14pt}

\usepackage{mwe} 

\usepackage{xcolor}

\let\origsection\section
\renewcommand\section[1]{\vspace{-1truemm}\origsection{#1}\vspace{-1truemm}}

\let\origsubsection\subsection
\renewcommand\subsection[1]{\vspace{-2truemm}\origsubsection{#1}\vspace{-1truemm}}

\let\origsubsubsection\subsubsection
\renewcommand\subsubsection[1]{\vspace{-2.5truemm}\origsubsubsection{#1}\vspace{-1.5truemm}}

\title{Tracking intermittent particles with self-learned visual features}

\name{\begin{tabular}{c}Raphael Reme$^{\star \dagger}$ \qquad Victor Piriou$^{\star}$\qquad Alison Hanson$^{\ddagger}$ \qquad Rafael Yuste$^{\ddagger}$\qquad Alasdair Newson$^{\dagger}$\\ \qquad Elsa Angelini$^{\dagger}$ \qquad
  Jean-Christophe Olivo-Marin$^{\star}$\qquad Thibault Lagache$^{\star}$\end{tabular}}

\address{\small$^{\star}$ Institut Pasteur, Université de Paris-Cité, CNRS UMR 3691, BioImage Analysis Unit F-75015 Paris, France\\
    \small$^{\dagger}$ LTCI, Telecom Paris, Institut Polytechnique de Paris, France\\
    \small$^{\ddagger}$ Department of Biological Sciences, Columbia University, New-York, U.S.A.\\
    \small Corresponding author: raphael.reme@pasteur.fr \vspace{-3truemm}
    }
    
\usepackage{eso-pic}
\AddToShipoutPictureBG{
    \AtPageUpperLeft{
        \hspace{\oddsidemargin}\hspace{1truein}
        \raisebox{-3\baselineskip}{
            \parbox{\textwidth}{
                \centering \color{gray} 
                \textcopyright~2023 IEEE. Personal use of this material is permitted.
                Permission from IEEE must be obtained for all other uses, in any current or future media,
                including reprinting/republishing this material for advertising or promotional purposes,
                creating new collective works, for resale or redistribution to servers or lists,
                or reuse of any copyrighted component of this work in other works.
                \href{https://doi.org/10.1109/ISBI53787.2023.10230664}{DOI:~10.1109/ISBI53787.2023.10230664}
            }
        }
    }
}

\begin{document}
\ninept
\maketitle
\begin{abstract}
In time-lapse fluorescence imaging, single-particle-tracking is a powerful tool to monitor the dynamics of objects of interest, and extract information about biological processes. However, tracked particles can be subject to occlusion and intermittent detectability. When these phenomena persist over a few frames, tracking algorithms tend to produce multiple tracklets for the same particle. In this work, we introduce self-supervised learning of visual features to compare tracked particles, and we exploit both visual and positional distances to robustly stitch tracklets representing the same particle. We demonstrate the performance of our stitching framework on time-lapse fluorescence sequences of \emph{Hydra Vulgaris} neurons. Results show high stitching precision, and reduction of errors made by previous algorithms on the same data by a factor of two.
\end{abstract}
\begin{keywords}
  Single Particle Tracking, Optimization, Deep Learning, Self-supervised Learning
\end{keywords}
\section{Introduction}
\label{sec:intro}
To study the dynamics of particles (e.g., molecules, pathogens, cells...) in fluorescence imaging, single particle tracking (SPT) is required to extract meaningful information for each tracked particle. SPT is usually divided into two different stages. First, particle detection and localization is performed on each time frame. Then, the detections are linked into coherent tracks.

The detection of particles' spots in biological images can be solved in different ways. Standard approaches typically involve multiple steps including noise reduction (e.g. Gaussian smoothing, wavelet denoising...), signal enhancement (e.g. top-hats, h-domes...) and thresholding \cite{smal2009quantitative}. Newer
algorithms using deep learning such as StarDist \cite{schmidt2018cell} or simple object detection approaches
such as faster-RCNN or Yolo \cite{ren2015faster, redmon2017yolo9000} have shown better performance,
but require manual annotation of spots for training.

Linking detections through time is complex when numerous interacting particles are involved. In addition, the detection step usually leads to missed and false detections due to poor signal-to-noise ratio (SNR). Therefore, elaborate tracking algorithms have been developed over the years to robustly link detections into coherent tracks for each particle. A first class of algorithms relies on global distance minimization (GDM) between consecutive frames detections \cite{sbalzarini2005feature,jaqaman2008robust}. But frame-to-frame linking is not sufficient, and solving the GDM problem
over all frames is infeasible due to memory and time limitations. Therefore, heuristic methods must be used.
In \cite{jaqaman2008robust} the frame-to-frame GDM is solved to produce tracks and GDM is reapplied globally to
merge and split tracks and resolve inherent issues of frame-to-frame linking.
Tracking can also be solved with probabilistic frameworks. For instance, using Kalman filters or Interacting Multiple Models
(IMM) with likelihood maximization \cite{1621229}. Multiple Hypothesis Tracking (MHT) heuristics can also be used. This keeps
several hypotheses at each frame, allowing the final linking decision to be made after seeing a few following frames \cite{chenouard2013multiple}.

Due to imaging conditions (e.g. low SNR, particles moving out-of-focus, occlusions...) or the biophysics of the imaged probe, that can switch into a non-detectable state (e.g. intermittency of Quantum Dots fluorescence \cite{bonneau2005single}, photo-switchable fluorophores \cite{rust2006sub}, calcium imaging of neurons \cite{smetters1999detecting}...) tracked particles can remain undetectable for a few frames. State-of-the-art tracking algorithms struggle to track such intermittent particles (that are undetectable for more than $\sim 5$ frames). Instead, they predict multiple partial tracks for the same particle, called \emph{tracklets}. Therefore, a post-processing step, referred to as \emph{tracklet stitching} is required to link tracklets of the same particles into one single track. In \cite{bonneau2005single}, local intensity was used to find a minimal path between tracklets. The minimal path method was specifically designed for well separated particles. Therefore, more versatile methods based on the computation of a tracklet-to-tracklet cost, followed by computation of optimal linking with GDM were subsequently developed \cite{jaqaman2008robust,Lagache2020.06.22.165696}. However, none of the cost used for tracklet stitching in these methods fully exploited visual features of the tracked particles.

In this paper, we address the tracklet stitching problem. Our main contribution is the incorporation of both visual and positional distances when building the tracklet-to-tracklet cost. We use GDM to find the optimal stitching of multiple tracklets into single tracks. To define a robust visual tracklet-to-tracklet cost, square image patches around spots' detections in tracklets are extracted. Patches are then embedded into discriminative features using a self-supervised trained convolutional neural network (CNN). Our definition of the positional cost follows previous works \cite{jaqaman2008robust,Lagache2020.06.22.165696} and integrates motion correction. Finally, we propose an algorithm that learns the optimal mixture of visual and positional cost from a few manually annotated links. We demonstrate the performance of our tracklet stitching framework on time-lapse fluorescence sequences of \emph{Hydra Vulgaris} neurons, reducing the errors make by previous state of the art algorithms by a factor of two.



\section{Method}

\begin{figure*}
  \centering
  \includegraphics[width=\linewidth]{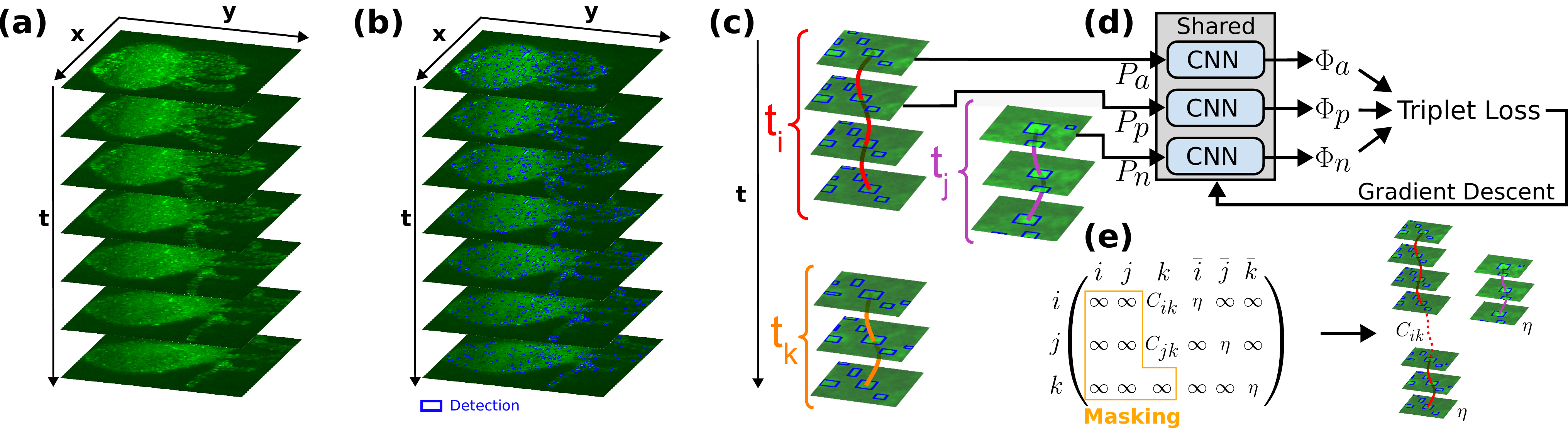}
  \vspace{-6truemm}
  \caption{Global overview of the method. \textbf{(a)} Input time-lapse sequence (fluorescence imaging). \textbf{(b)} Particles' spots are detected in each frame. \textbf{(c)} The detections are linked into tracklets. Here, $t_i$ and $t_k$ track the same particle. \textbf{(d)} Self-supervised learning of discriminative visual features for patches around spots' detections. Triplet loss is used to produce particle-specific visual feature clusters. \textbf{(e)} Learned visual features and particles' positions are used to define a hybrid tracklet-to-tracklet cost. Temporal masking of backward links reduces computational load and improves  stitching performance. A tracklet can remain non-linked, \textit{i.e} linked to a virtual tracklet ($\bar{i}, \bar{j}, \bar{k}$) with cost $\eta$. The minimal assignment cost gives the links between tracklets. Here, the minimal cost is $C_{ik} + 2\eta$ for linking $t_i$ to $t_k$ and $t_j$, $t_k$ to no one.}
  \label{fig:method}
  \vspace{-5truemm}
\end{figure*}

\subsection{Building tracklets}
Our method takes as input the results of a detection and tracking algorithm, which can be chosen freely. For the detection algorithm  (see Figure \ref{fig:method}.a-b), we chose to use B3-spline undecimated wavelet transform to extract spots as in \cite{OLIVOMARIN20021989}. This a very robust algorithm that does not require human labeling, but only manual definition of the spot scale and a detection threshold. Moreover, it is already implemented in Icy software (\textit{Spot Detector} plugin) \cite{de2012icy}. 
For the tracking algorithm, we used the probabilistic EMHT \cite{chenouard2013multiple}, a robust MHT heuristic which is able to handle false and missed detections on short time range. EMHT is also implemented in Icy (\textit{Spot Tracking} plugin).

For a given video, the tracking algorithm produces a set of $K$ tracklets $\mathcal{T} = \{t_i\}_{1 \le i \le K}$. Each tracklet can be defined as $t_i=\{x_i(\tau), \: \tau\in \Inter{\tau^s_i}{\tau^e_i}\}$, where $x_i(\tau) \in \R^2$ is the spatial position of the tracklet on frame $\tau$. The tracklet $t_i$ starts on frame $\tau_i^s \ge 1$ and ends on frame $\tau_i^e \le T$, where $T$ is the total number of frames of the time-lapse sequence.


\subsection{Tracklet stitching}
\label{sec:tracklet_stitching}

Among the computed tracklets, many might follow the same particles over time (e.g. $t_i$ and $t_k$ in Figure \ref{fig:method}.c) especially for intermittent particles. Therefore, a post-processing step is needed to link all the tracklets that refer to the same single particle, and create long-term, coherent particle trajectories. This task is referred to as \emph{tracklet stitching}. We denote a \emph{link} between two tracklets $t_i$ and $t_j$ as $(i \rightarrow j) \in \Inter{1}{K}^2$.

We address the tracklet stitching problem using GDM as previously done in \cite{Lagache2020.06.22.165696, jaqaman2008robust}. The main innovation of this paper is the construction of a hybrid tracklet-to-tracklet cost matrix $C \in \R^{K\times K}$ that combines (learned) visual features and position information. The optimal linear assignment problem is then solved with Jonker-Volgenant algorithm \cite{jonker1987shortest}.

Several types of information (e.g. position, visual, motion) can be used to compare tracklets, and the optimal combination for tracklet stitching is usually unknown. The following sections describe the construction of a novel hybrid cost matrix $C$ that integrates both visual and positional information to accurately stitch tracklets.

\subsubsection{Positional cost}

A simple way to define positional cost for linking two tracklets $t_i$, $t_j$ is to take the distance between the last position of the first tracklet and the first position of the second tracklet as in \cite{jaqaman2008robust}.
Let's assume that $\tau_i^e \le \tau_j^s$, then
\begin{equation}
  C^{\text{pos}}_{ij} = \norm{x_i(\tau_i^e) - x_j(\tau_j^s)}_2
\end{equation}
with $||.||_2$ the Euclidean norm. In some cases, particles are not independent. For example, when particle motion is induced by a common support, such as cells in tissues or molecules at the cell membrane. The motion of neighbor particles can then provide information on the trajectory of a given particle. With this in mind, the EMC2 algorithm \cite{Lagache2020.06.22.165696} predicts the forward and backward positions of particles using thin plate spline interpolation between positions of tracked neighbors. Let $\hat{x}^f_i(\tau)$ be the forward estimate of the tracked particle position after tracklet termination, \textit{i.e.} for frames $\tau \ge \tau_i^e$, and $\hat{x}^b_i(\tau)$ the backward estimate of the tracked particle position before tracklet initiation, \textit{i.e.} for frames $\tau \le \tau_i^s$. Then, assuming that $\tau_i^e \le \tau_j^s$, the positional cost defined in \cite{Lagache2020.06.22.165696} is given by
\begin{equation}
  C^{\text{EMC2}}_{ij} = \min_{\tau\in\Inter{\tau_i^e}{\tau_j^s}} \norm{\hat{x}^f_i(\tau) - \hat{x}^b_j(\tau)}_2
\end{equation}
\vspace{-1em} 
\subsubsection{Visual cost}

In addition to positional information, visual features can also be used to compare and define a cost between tracklets. Let assume that each tracklet $t_i$ at each frame $\tau$ can be characterized by visual features $\Phi_i^\tau \in \R^d$, with $d$ the dimension of the features. These features visually describe particles and are designed to discriminate between particles (see section \ref{sec:vis_features}). We can then define a visual cost for the link $(i \rightarrow j)$ as the smallest distance between the visual features of $t_i$ and $t_j$:
\begin{equation}
  C^{\text{vis}}_{ij} = \min_{\tau_i \in \Inter{\tau_i^s}{\tau_i^e}, \tau_j \in \Inter{\tau_j^s}{\tau_j^e}} \norm{\Phi_i^{\tau_i} - \Phi_j^{\tau_j}}_2
\end{equation}
\vspace{-1em} 
\subsubsection{Temporal masking}
To eliminate impossible or redundant links, we use temporal information, and define a boolean mask $M \in \{0,1\}^{K\times K}$ as $M_{ij} = (\tau_i^e \ge \tau_j^s)$. When $M_{ij} = 1$, the cost calculation is skipped, no link can be made between tracklet $t_i$ and tracklet $t_j$.

The mask simply states that the cost of $(i\rightarrow j)$ is computed only when $t_i$ ends before $t_j$ begins, allowing only forward temporal linking. The benefits are twofold: it improves computation time (less than half of the costs are computed) and it also improves the overall performance of the stitching because it removes the competition between backward and forward temporal links.

Indeed, with a particle tracked by three or more tracklets ($i \rightarrow j \rightarrow k$), the link $(j \rightarrow i)$ is incompatible with $(j \rightarrow k)$ and, without masking, the assignment algorithm would have to choose between the two.

\subsubsection{Final cost matrix}
\label{sec:cost_mixture}
Rather than choosing a single method to construct the cost matrix, we merge positional and visual information and define a more robust hybrid cost.

Let $(C^\text{pos}, C^\text{EMC2}, C^\text{vis}) = (C^m)_{1 \le m \le M}$ be the different candidate cost matrices. We first normalize each cost matrix in $\inter{0}{1}$: $\hat{C}^m = \frac{C^m - \min(C^m)}{\max(C^m) - \min(C^m)}$, before merging costs in a weighted average:
\begin{equation}
   C_{ij}(\alpha) = \vast\{
   \begin{aligned}
      & \infty                               &  & \text{ if } M_{ij} = \text{True} \\
      & \sum_{m=1}^M \alpha_m \hat{C}^m_{ij} &  & \text{ otherwise }
   \end{aligned}
\end{equation}
where $\alpha=[\alpha_1,\ldots,\alpha_M]$ are the weights associated to each positional and visual costs. These weights have to be tuned for each specific tracking problem, and we present in section \ref{sec:cost_mixture_learning} an optimization algorithm to learn a nearly optimal mixture of costs from few annotated tracklets links.

\subsubsection{Evaluation}

A video is manually annotated to have a set of $L$ true links $\mathcal{L} = \{(i \rightarrow j) \in \Inter{1}{K}^2\}$. The assignment problem is solved with the cost matrix $C(\alpha)$ and with cost limit $\eta \in \inter{0}{1}$. It predicts the set of links $\hat{\mathcal{L}}$, which is then compared to the annotation $\mathcal{L}$.

The choice of $\eta$ is particularly important. It represents the cost paid when linking a tracklet to no one (Figure \ref{fig:method}.e). Intuitively there are fewer, but more accurate predicted links when $\eta$ is closed to 0 (and higher recall, lower precision for $\eta$ close to 1).

We measure the Average Precision when the cost limit $\eta$ goes from $0$ to $1$, and find the best $\alpha$ by grid search.

\subsection{Learning visual features}
\label{sec:vis_features}
Visual comparison between tracked particles can increase the accuracy and robustness of tracklet stitching. But this requires defining a visual embedding of the particles. Previous works have used \textit{ad hoc} features (usually the shape, size or intensity of the tracked particle) \cite{sbalzarini2005feature, jaqaman2008robust, chenouard2008feature, chenouard2008improving}. Instead, we propose to use self-supervised learning to automatically learn the features that are relevant to this task.

We train a CNN encoder $\varphi$ to visually discriminate particles using only tracklet knowledge and thus requiring no human annotations. For each tracklet position $x_i(\tau)$, let $P_i^\tau$ be an $L\times L$ spatial square patch extracted from frame $\tau$ and centered on $x_i(\tau)$ (Figure \ref{fig:method}.c shows different patches around particles' spots for three tracklets). Each patch can be encoded into features $\Phi_i^\tau = \varphi(P_i^\tau) \in \R^d$. The training goal is to produce close features for patches from the same tracked particle, but distant features for different particles.

\subsubsection{Training}

Training is done with a single video and its tracklets. We extract all the patches for all the spots' detections within tracklets. Tracklets are divided into training and validation tracklets based on their position. We ensure that the training tracklets patches do not overlap with any of the validation ones. 70\% of tracklets are used for training.



The CNN is trained with the triplet loss \cite{wang2014learning} which is a standard loss for learning similarities and dissimilarities between images. Let $P_a$, $P_p$, $P_n$ be a triplet of patches (anchor, positive, negative), and $\Phi_a$, $\Phi_n$, $\Phi_p$ their respective features. The positive and anchor samples are from the same particle (and different than the negative one). Then the triplet loss with margin $m$ is
\begin{equation}
\label{eq:triplet_loss}
    L\left(\Phi_a, \Phi_p, \Phi_n\right) = \max\big(\norm{\Phi_a - \Phi_p} - \norm{\Phi_a - \Phi_n} + m, 0\big)
\end{equation}
Intuitively, the loss increases when the anchor features are closer to the negative ones than the positive ones.

To generate a triplet, we first sample an anchor tracklet $t_i$
and two random patches $P_a$ and $P_p$ from it (different frames of the same tracklet). Then we sample a tracklet $t_j$ which
overlaps temporally with $t_i$. This prevents $t_i$ and $t_j$ from representing the same particle. Finally, we sample
a random patch $P_n$ from $t_j$ (see Figure \ref{fig:method}.d).

\subsubsection{Validation}
We define a pretext and unsupervised task to validate the network with the validation tracklets. The visual aspect of a tracklet can be characterized by the average of its features $F_i = \frac{1}{\tau_i^e - \tau_i^s + 1} \sum_{\tau=\tau_i^s}^{\tau_i^e} \Phi_i^\tau$.

Then, we evaluate for each tracklet $t_i$ the capacity to retrieve the tracklet's features $\{\Phi_i^\tau\}_{\tau\in\Inter{\tau_i^s}{\tau_i^e}}$ in the closest features to the average features $F_i$. Our validation metric is the mean Average Precision (mAP) of this retrieval task.

\subsubsection{Implementation details}

We assume a dense particle distribution and set the patch size $L$ to be $\sim 3$ times the average closest spot distance ($L = 32$ pixels in our case) to include neighboring spots in each patch. This allows the model to learn particles' shapes and relationships. Patches intensities are normalized using the overall pixel mean and standard deviation in the training dataset.
We use a default ResNet18 \cite{he2016deep} architecture. After the last convolutional layer, adaptive
average pooling is used, and a final linear layer projects the features onto a 50-dimensional space.

A batch size of 512 is used, and learning rate is tuned on the validation set from $\inter{1e^{-5}}{1e^{-3}}$.
Weight decay is set to $0.01$ to avoid overfitting. We used a fixed version of Adam with weight decay:
AdamW \cite{loshchilov2018decoupled}. The momenta are set to $0.9$ and $0.999$.
The margin $m$ of the triplet loss is 1.

The CNN is trained for 20,000 steps with validation every 100 steps, the best network on validation is kept.
A single training lasts less than 2h on a A100 Nvidia GPU.

\subsection{Costs mixture learning}
\label{sec:cost_mixture_learning}

In section \ref{sec:tracklet_stitching}, (nearly) optimal weights $\alpha$ are found by grid search which require a complete annotation of tracklet links for at least hundreds of frames.

We present here an alternative to grid search using much fewer manually annotated links so that $\alpha$ can be easily tuned on new data. Let's assume to have annotated only a small subset of the links $\tilde{\mathcal{L}} \subseteq \mathcal{L}$ of cardinal $\tilde{L} \ll L$. Our approach then focuses on a single linear optimization problem.



For each manually-annotated link $(i \rightarrow j) \in \tilde{\mathcal{L}}$, the cost $C_{ij}(\alpha)$ should be smaller than the cost of any concurrent link.
Formally we define, for a given link $(i \rightarrow j)$, the set of concurrent links as:
\begin{equation}
  \overline{\mathcal{L}}(i\rightarrow j) = \left\{(p \rightarrow q)|\,p = i \text{ or } q = j\right\},
\end{equation}
in other words, all links that share a common tracklet with $(i \rightarrow j)$.

An ideal cost matrix should verify that for all annotated links in $\tilde{\mathcal{L}}$, the associated cost should be minimal, compared to the other concurrent links, that corresponds to the optimization problem 

\begin{equation}
    \min_{\alpha} \underset{(p \rightarrow q)\in \overline{\mathcal{L}}(i \rightarrow j)}{\max_{(i \rightarrow j) \in \tilde{\mathcal{L}}}}\left\{\begin{aligned}
    C_{ij}(\alpha) - C_{pq}(\alpha)
    \end{aligned}\right\}
\end{equation}
By introducing a margin $\epsilon$, we solve the equivalent linear minimization problem with a Simplex:

\begin{equation}
  \begin{aligned}
     & \text{min}_{\epsilon, \alpha}\ \ &  & \epsilon                                                                                                \\
     & \text{with}\                    &  & \alpha^\intercal 1_M = 1 \; \mathrm{and} \; \alpha \ge 0                                                            \\
     &                                 &  & \forall (i \rightarrow j) \in \tilde{\mathcal{L}},\; \forall (p \rightarrow q)\in \overline{\mathcal{L}}(i \rightarrow j),\; \\
     &                                 &  & C_{ij}(\alpha) \le C_{pq}(\alpha) + \epsilon
  \end{aligned}
  \label{eq:good_mixture}
\end{equation}


\subsubsection{Evaluation}

Among all the annotated links $\mathcal{L}$ (Section \ref{sec:tracklet_stitching}), we randomly sample a subset $\tilde{\mathcal{L}}$ of $\tilde{L}$ links and learn the weights $\alpha$ by solving the resulting optimization problem. We evaluate the resulting cost $C(\alpha)$ on all the links, as in section \ref{sec:tracklet_stitching}. We repeat this operation 50 times for each value of $\tilde{L}$ to account for the variability of the method due to link sampling.

\section{Results}


\begin{table}
  \begin{center}
    \begin{tabular}{l|c|c|c|c}
      Video         & \textbf{1} & \textbf{2} & \textbf{3} & \textbf{4} \\
      \hline
      Nb. Frames        & 251        & 401        & 401        & 1000       \\
      Marker type       & M1         & M1         & M2         & M2         \\
      Nb. tracklets & 784        & 1275       & 1763       & 4022       \\
      Median tracklet length & 27         & 19         & 39         & 68.5\\
      Nb. patches & 45k & 75k & 162k & 768k \\
      \hline
      mAP (Raw) &98.2\% & 96.6\% & 91.5\% & 77.9\%\\
      mAP (Trained) & \textbf{99.3\%}     & \textbf{98.4\%}     & \textbf{94.8\%}     & \textbf{97.6\%}     \\
      
    \end{tabular}
  \end{center}
  \vspace{-5truemm}
  \caption{Data and self-supervised visual training results on \emph{Hydra Vulgaris}.
    Marker types are GCaMP (M1) or TdTomato (M2). We compare the mean Average Precision (mAP) when using trained features versus \emph{raw} features (flatten images as features).
  }
  \vspace{-2truemm}
  \label{tab:features_learning}
\end{table}

We validate our method on time-lapse fluorescence sequences of \emph{Hydra Vulgaris} neurons with two different markers: calcium imaging with GCaMP (intermittent) or nuclear-targeting TdTomato. Sequences were acquired with a spinning disk confocal microscope.

On four different videos, tracklets were built using Icy software \cite{de2012icy}. We fully annotated the 784 tracklets on Video 1 (Table \ref{tab:features_learning}) with a total of 353 manual links, that we used to validate our tracklet stitching algorithm.

\subsection{Visual features learning}


Table \ref{tab:features_learning} summarizes the performance of our visual features learned by the CNN to retrieve tracklet's patches from their average features.

By construction, the task is easier with fewer and smaller tracklets, making comparison across video irrelevant. We demonstrate that we have good features by comparing with a raw features baseline on the same video, where we simply use the patches as the features (no encoder).

As patches are quite simple, raw features perform quite well. But on each video and each marker (GCaMP, TdTomato) the trained features outperforms the raw features. In addition, they are more robust as they adapt to the data: for instance in video 4, we observe a drop in performances using raw features that does not occur with trained features.



\subsection{Tracklet stitching}

\begin{table}
  \begin{center}
    \begin{tabular}{l c|c}
      Cost              & Masking & Links-AP \\
      \hline
      $C^{\text{pos}}$  & Yes $|$ No     & 49.1\% $|$ 22.6\%      \\
      $C^{\text{EMC2}}$ & Yes $|$ No  & 84.7\% $|$ 41.2\%    \\
      $C^{\text{vis}}$  & Yes $|$ No     & 64.6\% $|$ 26.8\%    \\
      \hline
      Best Mixture (Ours)           & Yes     & \textbf{92.0\%}\\
    \end{tabular}
  \end{center}
  \vspace{-5truemm}
  \caption{Tracklet links retrieval: Average precision (AP) with different cost matrix options.}
  \label{tab:tracklet_stitching}
\end{table}



\begin{figure}[t]
    \centering
    \includegraphics[width=\linewidth]{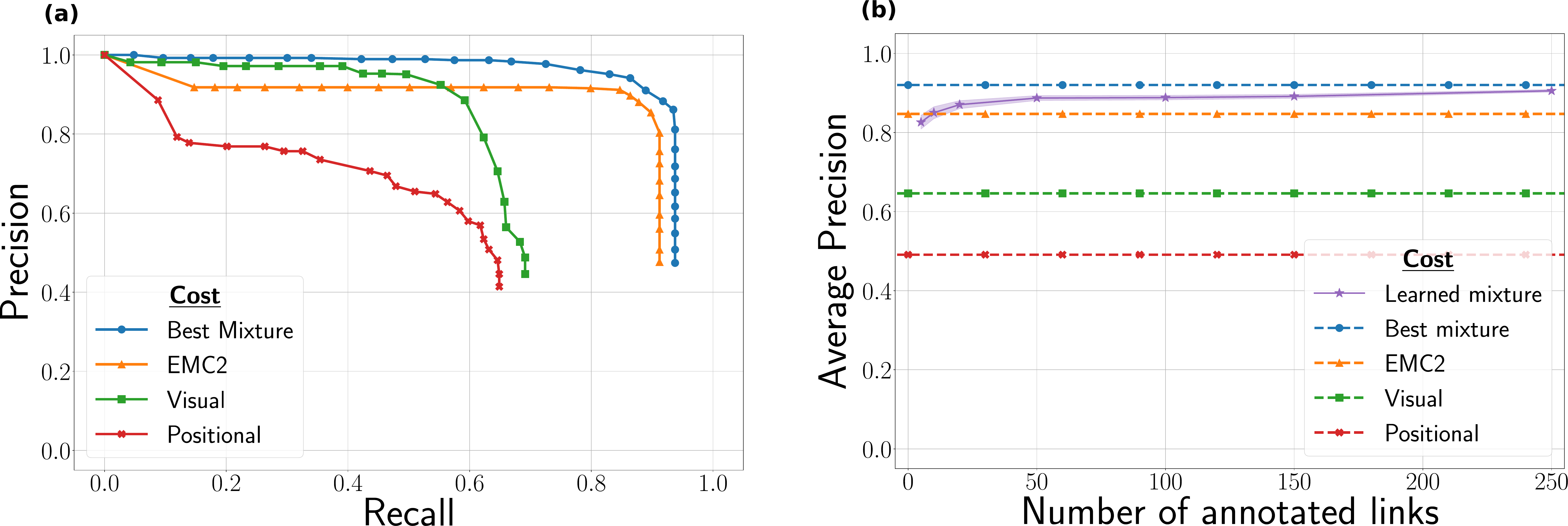}
    \vspace{-5truemm}
    \caption{\textbf{(a)} Precision-Recall curves for different cost matrix options. \textbf{(b)} Average precision versus number of annotated links when learning the costs mixture. (95\% confidence interval on 50 trials)}
    \vspace{-5truemm}
    \label{fig:results}
\end{figure}


Results for tracklet stitching with Video 1 are shown in Table \ref{tab:tracklet_stitching} and Figure \ref{fig:results}.a.
First, we observe that masking drastically improves the algorithm performance. For instance, the average precision is doubled for EMC2 cost with masking. We also point out that using a pure visual distance yields quite good performances, better than naive positional
distance. EMC2 is better than pure visual, which is expected for neurons tracking in \emph{Hydra}. Indeed, the visual information of spots is not rich enough compared to the positional information of neurons' neighbours.

Finally, errors are almost halved by our best costs mixture, reaching $92\%$ of AP. By grid search, the best weights found are $\alpha_{\text{EMC2}} = 0.9$, $\alpha_{\text{vis}} = 0.1$ and $\alpha_{\text{pos}} = 0.0$. This means that, with our data, the corrected positions is the most important distance for the robust stitching of tracklets, followed by visual features, while the raw positional cost is not used due to the important deformation of the animal. By defining the best threshold $\eta^\star$ as the one maximizing the f1 score, we find $\eta^\star=0.05$, reaching $90.1\%$ of f1 ($94.1\%$ precision and $86.4\%$ recall).
\vspace{-0.5em} 
\subsection{Learning costs mixture}



Figure \ref{fig:results}.b shows the average precision evolution with the number of annotated links $\tilde{L}$. With only a few annotations (20), our algorithm learns weights $\alpha$ yielding a better cost matrix than pure EMC2. If more links are annotated, it improves almost up to the best mixture that we found previously with grid search (using all the links).

\section{Conclusion}

In this paper, we proposed a post-processing method to robustly stitch tracklets of intermittent particles in SPT. Our algorithm merges different visual or positional tracklet-to-tracklet costs.

To build a robust visual cost, we introduced self supervised-learning to learn relevant embeddings for tracklet patches. We used a standard ResNet architecture for natural images, but in many biological applications, spots' features are not as rich as when tracking cars or people. Thus, we believe that smaller models could improve performances (and at least computational time). Moreover, we trained a single model for each video, but using a cross-video model should improve robustness and the ability to transfer to new videos without retraining the model.

Nonetheless, our method has demonstrated state-of-the-art performance on neuronal cells tracking in fluorescence calcium imaging. We also proposed an optimization algorithm to select the best costs mixture given very few annotations. Being able to find the best mixture with only a few annotated links is particularly encouraging for the versatility of our method and its extension to new data with very few hyperparameters to define.

\section{Compliance with ethical standards}
This data used in this study was obtained in line with the principles of the Declaration of Helsinki and approved by Columbia University.

\section{Acknowledgments}
This work is supported by the Institut Pasteur and France-BioImaging Infrastructure (ANR-10-INBS-04). R.R and T.L. are supported by the ANR (ANR-21-CE45-0020-01 REBIRTH). R.Y. is supported by the NSF (2203119) and Vannevar Bush Faculty Award (ONR N000142012828). A.H. is supported by T32MH018870, Leon Levy Fellowship in Neuroscience, and K99NS127851-01.

None of the authors declare to have a financial conflict of interest in the results of this study

\bibliographystyle{IEEEbib_short}
\bibliography{refs}
\vfill
\pagebreak

\end{document}